\documentclass[twocolumn,showpacs,preprintnumbers,amsmath,amssymb]{revtex4}
\usepackage{epsf}
\usepackage{rotating}

\usepackage{graphicx,epsfig}
\usepackage{dcolumn}
\usepackage{bm}
\usepackage{xcolor}

\def\cF{{\cal F}}

\newcommand{\req}[1]{Eq.~(\ref{#1})}
\newcommand{\avg}[1]{\langle #1\rangle}

\newcommand{\fig}[1]{Fig.~\ref{#1}}

\DeclareMathOperator*{\argmax}{{\rm{argmax}}}	

\makeatletter
\newcommand{\thickhline}{%
    \noalign {\ifnum 0=`}\fi \hrule height 1pt
    \futurelet \reserved@a \@xhline
}
\newcolumntype{"}{@{\hskip\tabcolsep\vrule width 2pt\hskip\tabcolsep}}
\makeatother

\begin{document}


\title{Evolving Powergrids in Self-Organized Criticality: An analogy with Sandpile and Earthquakes}
\author{Ho Fai Po$^1$, Chi Ho Yeung$^{1}$\footnote{chyeung@eduhk.hk}, An Zeng$^2$, K. Y. Michael Wong$^3$}
\affiliation{$^1$Department of Science and Environmental Studies, The Education University of Hong Kong, 10 Lo Ping Road, Taipo, Hong Kong, \\
$^2$School of Systems Science, Beijing Normal University, Beijing 100875, People's Republic of China, \\
$^3$Department of Physics, The Hong Kong University of Science and Technology, Clear Water Bay, Hong Kong}

\date{\today}

\begin{abstract}
The stability of powergrid is crucial since its disruption affects systems ranging from street lightings to hospital life-support systems. Nevertheless, large blackouts are inevitable if powergrids are in the state of self-organized criticality (SOC). In this paper, we introduce a simple model of evolving powergrid and establish its connection with the sandpile model, i.e. a prototype of SOC, and earthquakes, i.e. a system considered to be in SOC. Various aspects are examined, including the power-law distribution of blackout magnitudes, their inter-event waiting time, the predictability of large blackouts, as well as the spatial-temporal rescaling of blackout data. We verified our observations on simulated networks as well as the IEEE 118-bus system, and show that both simulated and empirical blackout waiting times can be rescaled in space and time similarly to those observed between earthquakes. Finally, we suggested proactive maintenance strategies to drive the powergrids away from SOC to suppress large blackouts.
\end{abstract}

\maketitle


\section{Introduction}

Self-organized criticality (SOC) corresponds to the mechanism a system self-organizes itself to achieve a critical state between different phases in the long run. It was first proposed by Bak et al in their seminal papers~\cite{bak87, bak88} as the mechanism underlying the characteristic $1/f$ power-spectrum~\cite{press87}, self-similar fractal structures~\cite{mandelbrot82} as well as power-laws~\cite{markovic14}, commonly observed in many different natural and artificial systems, across an extensive temporal and spatial scale~\cite{bak89}. Although alternative approaches have been suggested to produce similar phenomena~\cite{falconer04, sornette06, markovic14}, SOC was considered by some researchers to be the explanation for universal complexity in nature~\cite{bak96}, although this view is under strong debate~\cite{watkins16}. 

Despite the strong debate, SOC has triggered tremendous interest in various areas including physics, biology, astronomy, geology and economics. Especially, many systems in these areas exhibit a power-law distribution of fluctuations, spanning from minimal disturbances to system-wide extreme events~\cite{bak91}. These observations are consistent with those observed in the sandpile model~\cite{bak87, bak88}, a prototype of SOC, in which the addition of a single sand grain may lead to avalanches of any magnitude following a power-law distribution. These similarities have made sandpile model (as well as SOC) a good analogy for explaining extreme events in the other systems, ranging from earthquakes~\cite{bak02} and forest fires~\cite{drossel92} on the earth to solar flares on the sun~\cite{lu91}; from mass extinction events in ancient nature~\cite{bak93} to stock market crashes in modern society~\cite{stauffer99}.

In this paper, we will examine another type of extreme event which significantly impacts our daily life -- large blackouts caused by cascading failures in powergrids~\cite{buldyrev10}. As an example of their negative impact, the large blackout in August, 2003 in USA contributed to 90 deaths~\cite{anderson12} and was estimated to cost $\$$6.4 billion~\cite{anderson03}. Similar to other extreme events, blackouts were shown to follow power-law size distributions~\cite{carreras04} and SOC has been suggested to be the underlying mechanism~\cite{dobson07}. In particular, the continual effort to satisfy the increasing consumption by upgrading the powergrids incrementally may have put it at a critical loading marginally before the emergence of the phase with frequent blackouts, i.e. the system is in SOC~\cite{dobson07}. A model to demonstrate this idea was recently introduced in~\cite{hoffmann14}. Nevertheless, most of these studies only focus on the power-law cascade size distributions, while other phenomena and consequences of powergrids as SOC systems are not examined.

The goal of this paper is to introduce a simple model of \emph{an evolving powergrid}, i.e. a powergrid which serves a group of nodes with increasing energy demand, and upgrades its capacity everytime after an overloading failure. We will focus on the evolution of the powergrids beyond the extensively studied individual cascading failures~\cite{motter02}. In addition to the cascade size distribution, we will examine the distribution of waiting time between blackouts, the predictability of large blackouts as well as the spatial-temporal rescaling of blackout data~\cite{bak02, corral04}. We aim to show the connection between our model and (i) real powergrids, (ii) the sandpile model and hence SOC, and (iii) earthquakes (a potential candidate of SOC~\cite{bak02}); indirect connections between these three systems are then established. We tested our model on a real powergrid topology of the IEEE 118-bus system and obtained similar results. We also showed that real blackout data can be rescaled in space and time as we observed for earthquakes~\cite{corral04}. Finally, we examined various maintenance strategies and their ability to suppress large blackouts by driving the system away from SOC.

\section{Model}

We will introduce a model of evolving powergrid and shows its connection with the sandpile model~\cite{bak87, bak88}. Specifically, we consider a network with $N$ nodes, labeled by $i=1,\dots ,N$, where each node $i$ is connected with $k_i$ neighbors. Initially, a load $L_i$ is allocated to node $i$. A fraction $r$ of the nodes are considered as \emph{power stations} and are assigned with a negative infinite load. The rest of the nodes are considered as \emph{energy consumers} and are assigned a random positive load, which follows a normal distribution with a positive mean, truncated by discarding the negative side. Each node $i$ with a positive $L_i$ is required to acquire a sufficient amount of power from the power stations to satisfy its demand, i.e. 
\begin{align}
\sum_j A_{ji} I_{ji} = L_i,
\end{align}
where $A_{ji}$ is the matrix element of the $N\times N$ adjacency matrix ${\cal A}$, such that  $A_{ij}=1$ if nodes $i$ and $j$ are connected and otherwise, $A_{ij}=0$; the current $I_{ji}$ is the power flow from node $j$ to node $i$. We will implement the model on square lattices and a real powergrid topology. 

To satisfy the demand of the consumer nodes, power is transferred to them from the power plants via the powergrid, composed of $M$ transmission lines called \emph{links}. We adopt the direct current (DC) approximation and assume that currents on the links minimize the resistive power $\sum_{ij}A_{ij}I_{ij}^2 R_{ij}$ in transportation, where $R_{ij}$ is the resistance on link $(ij)$. To simplify the model, we assume that $R_{ij}=1$ for all links, but the results can be easily generalized to the case with $R_{ij}\neq 1$. As shown in~\cite{wong06}, the optimal current can be computed by solving the potential of each nodes, denoted as $v_i(t)$ for node $i$, given by \cite{wong06}
\begin{align}
\label{eq_v}
v_i(t) = \max\left[\frac{1}{k_i}\left(\sum_j A_{ij} v_j(t) + L_i(t)\right),0\right].
\end{align}
Since \req{eq_v} is a set of coupled equations involving a node and its neighbors, one can solve for $v$'s for all nodes by iterating the equations on the network until convergence~\cite{wong06}. The optimal current $I_{ij} (t)$ from node $j$ to node $i$ on a link $(ij)$ is then given by $I_{ij} (t)=v_i (t)-v_j (t)$, and $I_{ij} (t)=-I_{ji} (t)$.

Since links are transmission lines which may break when overloaded, we define the capacity of link $(ij)$ to be $C_{ij}(t)$ and consider that the link fails and no longer transfers electricity if the current $|I_{ij}|>|C_{ij}|$. Initially at time $t=0$, the capacity $C_{ij}(0)$ of a link $(ij)$ is defined as
\begin{align}
C_{ij}(0)=(1+\alpha)I_{ij}(0),
\end{align}
where $I_{ij}(0)$ is the optimal current of the link at time $t=0$, while $\alpha$ is the ratio of excess capacity installed in the network, which we call the \emph{redundancy ratio}. The capacity plays a role similar to the threshold in the sandpile model beyond which sands topple to the neighboring sites. Unlike the sandpile model, the capacity in the powergrid evolves with time due to repairs after a failure, given by 
\begin{align}
\label{eq_repair}
C_{ij}(t+1) = (1+\alpha){\tilde I}_{ij}(t),
\end{align}
where ${\tilde I}_{ij}(t)$ is the current which breaks the link. Capacities are heterogeneous across the network, which are different from the homogeneous thresholds in other SOC models such as the sandpile model, the OFC earthquake model~\cite{olami92} and the powergrid model in \cite{hoffmann14}. Yet, as we will see, this heterogeneity does not destroy criticality in our model, unlike that observed in the OFC model~\cite{janosi93}.

\begin{figure*}
\centering
\begin{tabular*}{0.95\linewidth}{ccccccc}
\multicolumn{1}{l}{(a)} & & \multicolumn{1}{l}{(b)} & & \multicolumn{1}{l}{(c)} & & \multicolumn{1}{l}{(d)} \\
\epsfig{figure=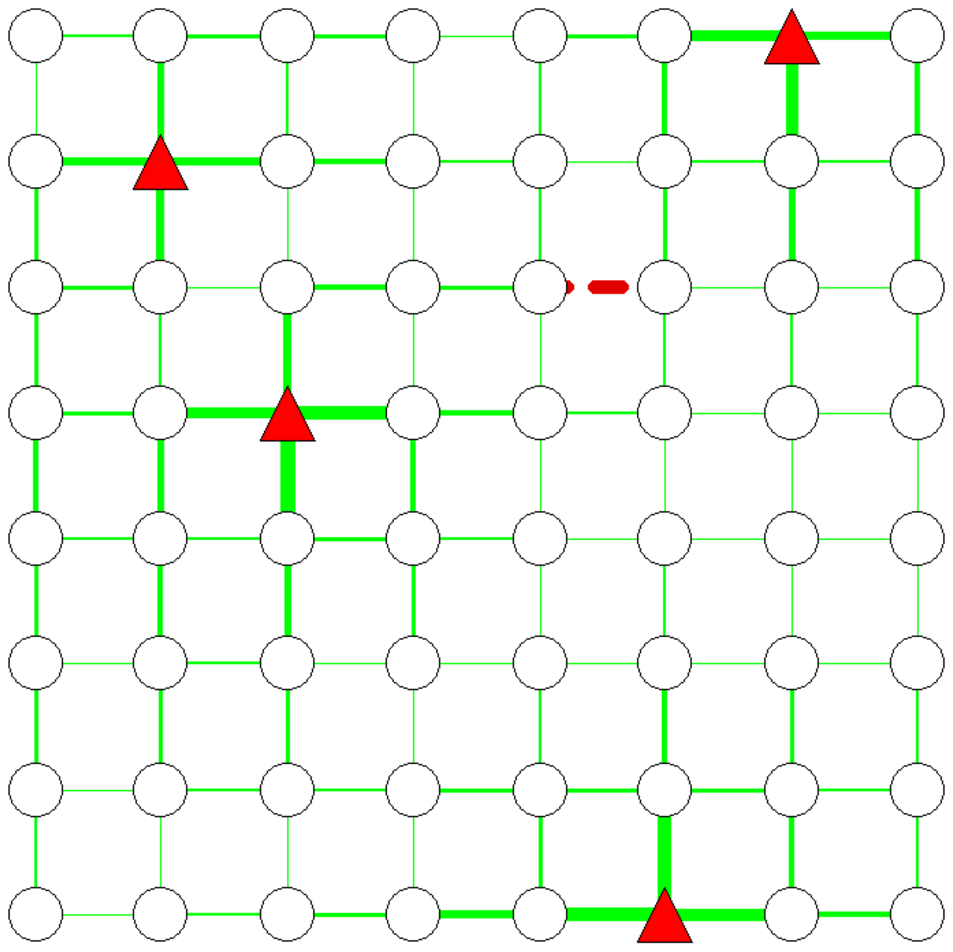, width=0.2\linewidth} & \hspace{0.02\linewidth} &
\epsfig{figure=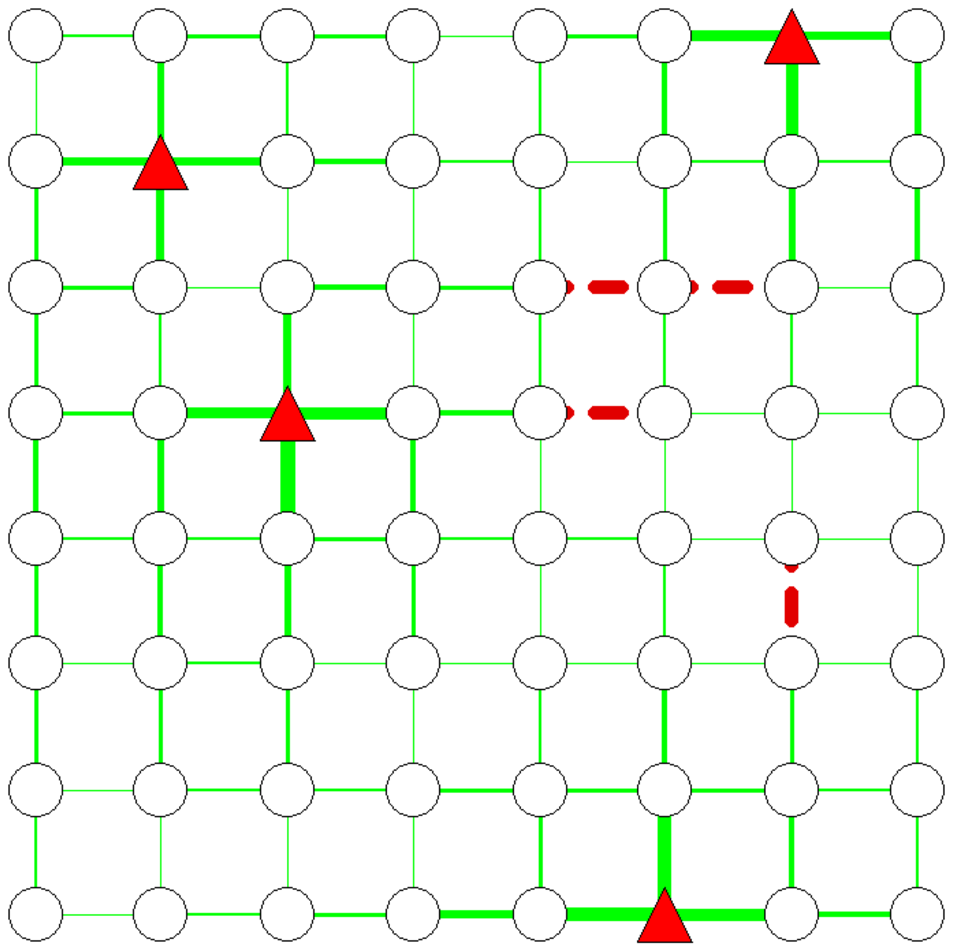, width=0.2\linewidth} & $\cdots$ &
\epsfig{figure=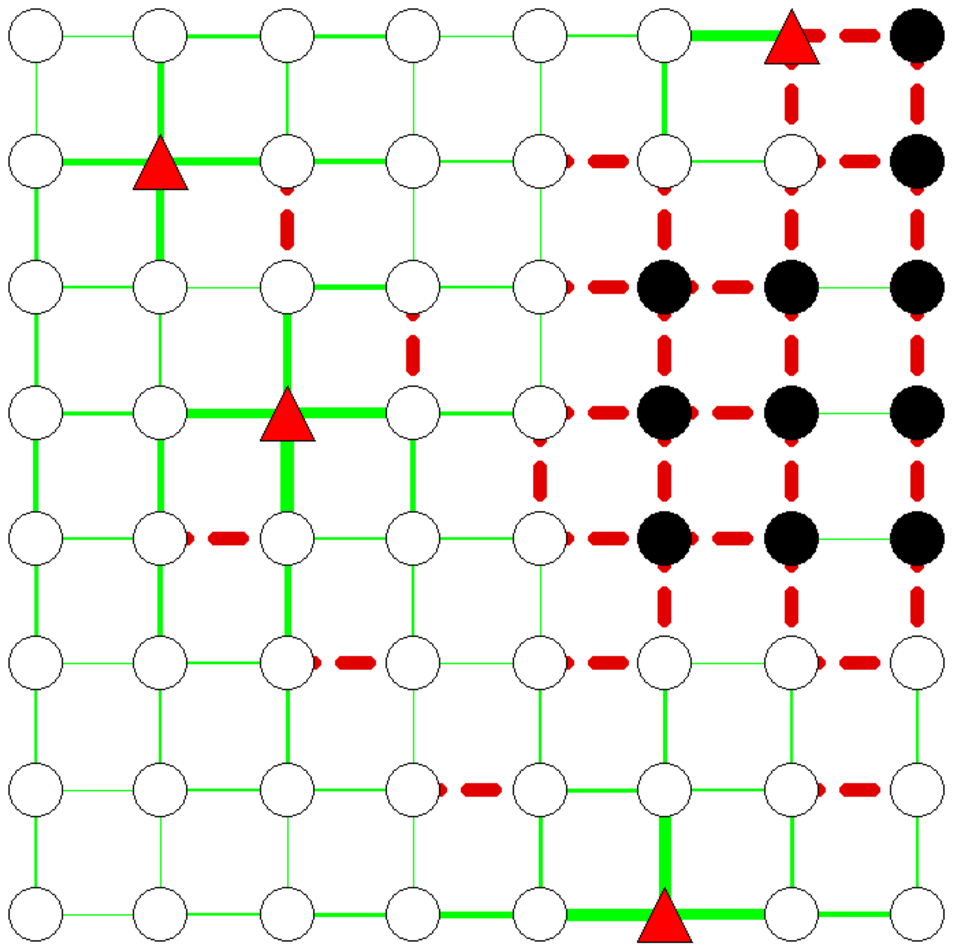, width=0.2\linewidth} & $\cdots$ &
\epsfig{figure=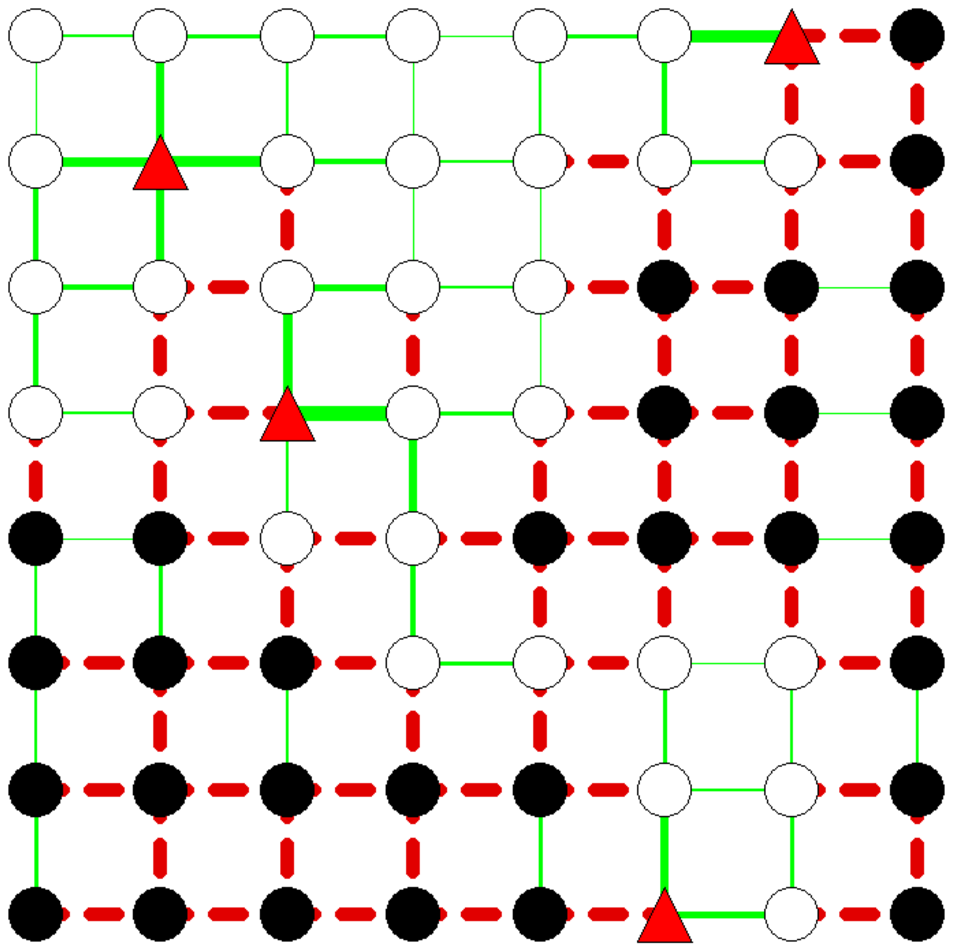, width=0.2\linewidth} \\
Cascade Step=1 & & Cascade Step=2 & & Cascade Step=4 & & Cascade Step=6
\end{tabular*}
\caption{
The dynamics of a cascading failure, or a \emph{blackout}, in our model. Power stations and consumer nodes are denoted as triangles and circles, respectively. Shaded circles correspond to consumer nodes which are disconnected from the power stations due to the failed links. (a) A link in the network is broken after a demand increment. (b) and (c) Power transmitted by the broken link is shared by the other links, which overload the other links to the same power station and render it isolated; the load in the network are now shared by the remaining power stations, leading to failures in areas away from the initially failed link. (d) The end of the cascade and all the failed links in this single blackout.
}
\label{fig_break_1}
\end{figure*}

To model the increase in the demand for electricity, we randomly pick a consumer node at time $t$ and increase its demand by
\begin{align}
L_i(t) = L_i(t-1) + \beta\avg{L_i(t-1)},
\end{align}
where $\avg{L_i(t)}=\sum_{j=1}^{N}L_j(t)/N$ and $\beta$ is the \emph{demand increment ratio}. This is analogous to the addition of sands to the sandpile model. After each demand increment, the new optimal currents are computed by iterating \req{eq_v}, and links with currents exceeding the capacity are considered broken and removed from the powergrid. Currents are then re-computed and new broken links are identified, leading to a \emph{cascading failure} as shown in \fig{fig_break_1}(a) and (b); some nodes may become disconnected from the powergrid as shown in \fig{fig_break_1}(c). But unlike the sandpile model, a failed link may affect links other than the nearest ones, as shown in the transition from \fig{fig_break_1}(a) through (c). The process repeats until no extra link breaks, and all failed links are then repaired according to \req{eq_repair}. The whole cascading failure and repairs are finished within one time step before the next demand increment, which corresponds to a fast relaxation dynamics intercepting a slow driving dynamics, an essential mechanism in many SOC systems~\cite{bak96}. As the procedures repeat, a series of cascading failures is observed. In this paper, we call a failure \emph{a blackout} regardless of its size.

To keep the model simple, we set $\alpha=\beta$ such that the rates of load and capacity increment are equal. In this case, the model only captures the most essential components of the system to illustrate its relevance with SOC. While capacities and loads monotonically increase, we rescale all loads and capacities by the same factor after a number of simulation time steps to keep all the variables finite; the dynamics between the variables does not change since all equations involving loads and capacities are linear. Since we will focus on the SOC phenomenon in this paper, the analyses of the system behaviors with different $\alpha$ are given in the \emph{Appendix C}.

\section{Results}

\subsection{Self-organized criticality in the model powergrid}

\begin{figure}
\centerline{\epsfig{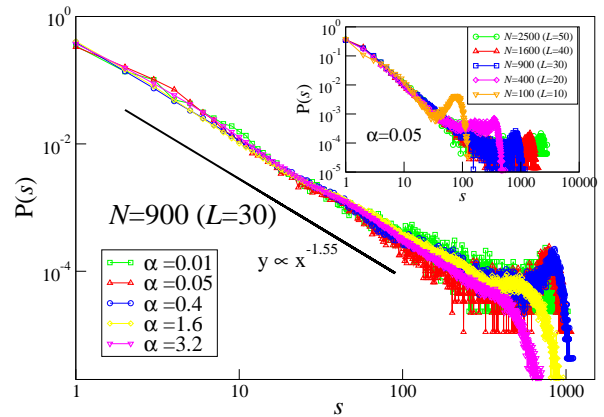}}
\caption{
The distribution of cascade size, i.e. the total number of broken links in each failure, obtained on $L\times L$ square lattices with $L=30$ and $N=L^2=900$ nodes, with $5\%$ of the nodes to be power station. The statistics is obtained for at least $N\times 2 \times 10^3$ updates after an equilibration of $N\times 10^2$ updates. Inset: The distributions of cascade size with different system size $N$ with $\alpha=0.05$.
}
\label{fig_sizeDis}
\end{figure}

To show that our powergrid model exhibits the characteristics of a critical state, we first examine the distribution of cascade size, measured as the total number of broken links in a failure. As shown in \fig{fig_sizeDis}, the cascade size follows a power-law distribution similar to the empirical results~\cite{dobson07}. Although different tails are observed for cases with different model parameter $\alpha$, we remark that a universal power-law exponent is observed with $\alpha$ spanning two order of magnitudes from $0.01$ to $3.2$, which emerges from self-organization without fine-tuning, suggesting universality independent of model parameters. On the other hand, as shown in the inset of \fig{fig_sizeDis}, the cascade size distributions for different system sizes also follow power-laws with the same exponent where the cutoff increases with system size, suggesting scale invariance typical of critical systems.

\begin{figure}
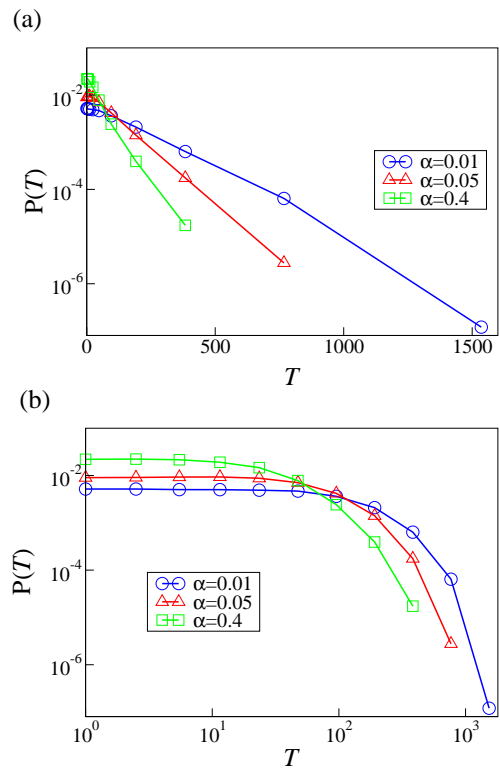

\centerline{
\epsfig{figure=linlogintereventtime.eps, width=0.75\linewidth} 
}

\centerline{
\epsfig{figure=loglogintereventtime.eps, width=0.75\linewidth}
}
\caption{
The distributions of waiting time between consecutive blackouts of size $s\ge 1$, obtained on an instance of $L\times L$ square lattice with $L=30$ and $N=L^2=900$ nodes, with $5\%$ of the nodes to be power stations, in (a) semi-log plot and (b) log-log plot. The statistics is obtained for $N\times 10^4$ updates after an equilibration of $N\times 10^2$ updates. To better show the form of the distributions, log-bins are used to obtain the distributions.
}
\label{fig_intereventtime}
\end{figure}

To further examine the criticality of our powergrid model, we compute the distribution of waiting time $T$ between consecutive failures. Since avalanches in SOC systems are unpredictable, their waiting time distributions should follow an exponential decay, though it is not common to use merely the exponential waiting time distribution to justify SOC~\cite{sanchez02, boffetta99, aschwanden11}. As shown in the semi-log plot in \fig{fig_intereventtime}(a), the waiting time distributions for the cases with different $\alpha$ follow an exponential decay. The distributions in log-log plot in \fig{fig_intereventtime}(b) show that they resemble the empirical results observed in real powergrids~\cite{dobson07}, and earthquakes~\cite{corral04} except a more prominent power-law at the small time scale contributed by the correlated aftershocks after a main shock described by the \emph{Omori's Law}~\cite{omori1894}. After-blackouts do not occur in our model and instead, failures occur less frequently after a large blackout since a large number of links were repaired simultaneously.

To examine the predictability of the occurrence and the size of blackouts, we analyze the sequence of waiting times and cascade sizes by \emph{detrended fluctuation analysis} (DFA); detailed description of DFA is found in the \emph{Appendix A}. As shown in \fig{fig_DFAInter} and \fig{fig_DFACas}, the behavior of the sequence of waiting times and cascade sizes is analogous to that of the corresponding reshuffled sequence when a small number of consecutive blackouts are considered. This implies that the exact behavior of blackout occurrence is unpredictable in the short time scale. On the other hand, the behavior of the sequences deviates from that of the reshuffled one when a large number of consecutive blackouts are considered; it implies that the emergence of blackouts become more predictable in the long run, e.g. a large blackout or a long waiting time emerges when a sufficiently long sequence of blackouts is considered. Nevertheless, the larger the system size, the longer the time scale which is consistent with the reshuffled results, implying a lower predictability in larger systems.

\begin{figure}
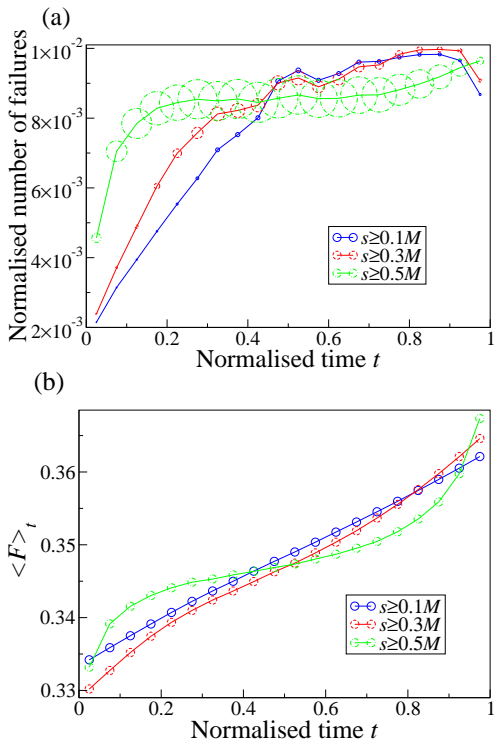

\centerline{
\epsfig{figure=NormalisedSize.eps, width=0.75\linewidth} 
}
\centerline{
\epsfig{figure=NormalisedIC.eps, width=0.75\linewidth}
}
\caption{
(a) We divide the duration between consecutive pairs of large blackouts into 20 divisions and compute the failure rate within each division averaged over different pairs of large blackouts with  size $s\ge 0.1M$, $0.3M$ and $0.5M$. The size of the nodes corresponds to the average blackout size in that particular division.
The results are averaged over 27 realizations of $L\times L$ square lattice with $L=30$ and $N=L^2=900$ nodes, with $5\%$ of the nodes being power stations. The statistics is obtained for $N\times 10^4$ updates after an equilibration of $N\times 10^2$ updates. 
(b) The corresponding average network load factor $F$ given by \req{eq_loadF}.
}
\label{fig_normalized}
\end{figure}

Stimulated by the DFA results, we examine the presence of blackout precursors. We first divide the time between consecutive pairs of large blackouts into a fixed number of divisions, and compute statistics within each division to identify potential trends. As shown in \fig{fig_normalized}(a), the failure rate increases and drops before the emergence of medium blackouts with size $s\ge 0.1M$ and $s\ge 0.3M$, where $M$ is the total number of links, which may signal an insufficient relaxation of energy before the blackouts. On the other hand, the failure rate increases continuously before blackouts of size $s\ge 0.5M$, which may be caused by a sequence of medium-sized blackouts (suggested by the large average blackout size) relaxing energy incompletely, and finally triggered by a sequence of small but more frequent blackouts, indicating a close-to-critical state of the system.

Nevertheless, \fig{fig_normalized}(a) shows the average failure rate while individual events may be unpredictable by just examining the preceding sequence of blackouts. Prediction may instead be possible if the microscopic information of the system is available. To examine this predictability, we define the \emph{load factor} of link $(ij)$ to be $F_{(ij)}=|I_{ij}|/C_{ij}$, and compute the \emph{average network load factor} $F$ given by
\begin{align}
\label{eq_loadF}
F = \frac{1}{M}\sum_{(ij)}F_{ij} = \frac{1}{M}\sum_{(ij)}\frac{|I_{ij}|}{C_{ij}}.
\end{align} 
A small value of $F$ implies a large amount of redundant capacity and failures are less likely. As shown in \fig{fig_normalized}(b), $F$ increases between two large blackouts, and a high $F$ may be a signal for coming large failures. These results suggest that prediction is difficult given only macroscopic information, e.g. the sequence of failures, but may be possible if microscopic information in the system is available. This is analogous to the predictability of earthquakes suggested in~\cite{bernard01}.

\begin{figure*}
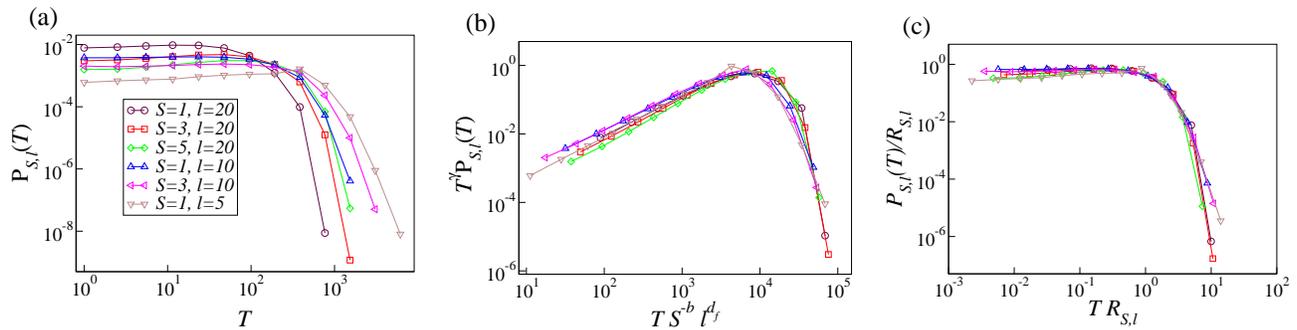

\centerline{
\epsfig{figure=Original.eps, width=0.3\linewidth} \quad
\epsfig{figure=BasicRescale.eps, width=0.3\linewidth} \quad
\epsfig{figure=Rate.eps, width=0.3\linewidth}
}
\caption{
(a) The distribution $P_{S,l}(T)$ of waiting time $T$ between consecutive failures with size $s\ge S$ in a $l\times l$ partition of the network, for different values of $S$ and $l$. (b) The data collapse of all the distributions in (a) by plotting $T^\gamma P_{S,l}(T)$ as a function of $x=TS^{-b}l^{d_f}$, suggested in~\cite{bak02} for earthquakes, with $\gamma =1.07$, $b=0.54$ and $d_f=1.5$. (c) The data collapse of all the distributions in (a) by plotting $R_{S,l}^{-1}P_{S,l}(T)$ as a function of the rescaled waiting time $x=R_{S,l}T$, suggested in \cite{corral04} for earthquakes, where $R_{S,l}$ is the average event rate with size $s\ge S$ in the partition of size $l$. The results are obtained on $L\times L$ square lattices with $L=20$ and $N=L^2 =400$ nodes, with $5\%$ of the nodes to be power stations. The statistics is obtained for $N\times3 \times 10^5$ updates after an equilibration of $N\times 10^2$ updates. 
}
\label{fig_rescale}
\end{figure*}

As suggested by \cite{bak87, bak88, watkins16}, spatio-temporal correlation is an important feature of SOC systems since it connects the dynamical self-organization with spatial criticality. To examine this correlation, we consider a $l\times l$ partition (details are shown in Appendix B), and define $P_{S,l}(T)$ to be the waiting time distribution within the partition between blackouts with size $s>S$. As shown in \fig{fig_rescale}(a), an increase in the \emph{cutoff size} $S$ (or a decrease in partition size $l$) makes the distributions wider, compared to the original $P_{S=1, l=20}(T)$. Hence, it is interesting to examine if the statistics of large failures in large partitions are equivalent to those of small failures in small partitions. We follow the analyses of earthquakes in~\cite{bak02}, and plot the rescaled distributions $T^\gamma P_{S,l}(T)$ as a function of the rescaled waiting time $x=TS^{-b}l^{d_f}$ in \fig{fig_rescale}(b), such that $P_{S,l}(T)$ with various $S$ and $l$ collapse onto the same function $f(x)$, i.e.
\begin{align}
\label{eq_collapse}
T^\gamma P_{S,l}(T) = f(TS^{-b}l^{d_f}).
\end{align}
As suggested in \cite{bak02}, $\gamma$, $b$ and $d_f$ corresponds to the exponents of the Omori's Law, the exponent of the cumulative distribution of shock magnitude (i.e. the exponent of Gutenberg-Richter Law~\cite{gutenberg54}), and the fractal dimension of earthquakes, respectively. In our case, $\gamma =1.07$, $b=0.54$ and $d_f=1.5$, and the value $-b-1$ is consistent with the exponent $\approx -1.54$ in the blackout size distribution in \fig{fig_sizeDis}, consistent with relation suggested by Bak \emph{et al} for empirical earthquake data~\cite{bak02}. This data collapse in \fig{fig_rescale}(b) suggests that waiting times between blackouts in our model can be rescaled in spatial and temporal dimension, again suggesting universality, and is similar to that observed for earthquakes.

To further examine the universality of blackout waiting time in our model, we follow an alternative form of rescaling proposed by~\cite{corral04}. By drawing again an analogy with earthquakes, the distributions $P_{S,l}(T)$ with various $S$ and $l$ for blackouts collapse onto the function $g(x)$ as shown in \fig{fig_rescale}(c), i.e.
\begin{align}
\label{rescale_rate}
R_{S,l}^{-1}P_{S,l}(T) = g(R_{S,l}T),
\end{align}
where $R_{S,l}$ is the average blackout frequency in the partition. The two data collapses in \fig{fig_rescale}(b) and (c) may suggest that scale invariance and universality emerge in our model of powergrid, resembling those observed for earthquakes~\cite{bak02, corral04}. All the above results suggest that our powergrid model results in a realistic blackout statistics as well as the characteristics of self-organized criticality.

\section{SOC on real powergrids}

\begin{figure}
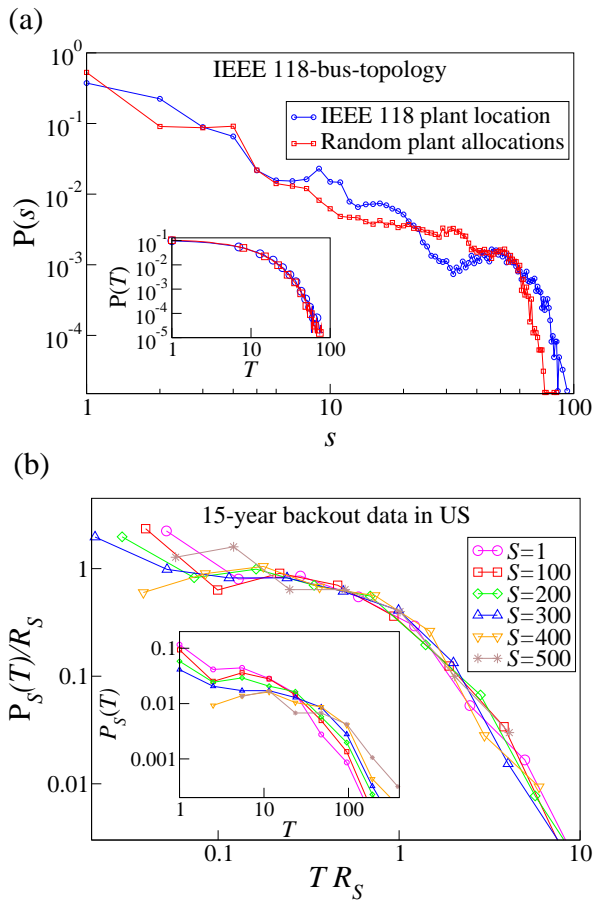

\centerline{
\epsfig{figure = 118PowerLaw.eps, width=0.9\linewidth}
}
\centerline{
\epsfig{figure = RealData.eps, width=0.9\linewidth} 
}
\caption{
(a) The distributions of blackout size, obtained on the IEEE 118-bus system with $N=118$, 105 consumer nodes, $15$ power stations and 186 links. Blue circles and red squares correspond to the cases where power stations are located at the original location and random location, respectively. The statistics is obtained for $N\times10^4$ updates after an equilibrium of $N\times10^2$ updates. Inset: The waiting time distributions of the two cases. (b) The rescaled waiting time distributions of the 15-year data of power outages across the United State~\cite{realdata}, $R_{S}^{-1}P_{S}(T)$ as a function of the rescaled waiting times $x=R_{S}T$, where $R_{S}$ is the average occurrence rate of failure of size greater than $S$. Inset: The waiting time distributions of the 15-years data of power outages across the United State with failure of size $s>S$.}
\label{fig_realdata}
\end{figure}

To examine the robustness of the SOC phenomenon against network topology, we implement our model on the IEEE 118-bus system, which represents a portion of the American Electric Power System in the Midwestern US as of December, 1962 \cite{118system}. Before the simulation starts, we assign a load $L_i$ to each consumer node $i$ according to its load given in \cite{118system}; power stations are assigned with an infinite amount of resources. The system then follows the same dynamics described in our model. As shown in \fig{fig_realdata}(a), the cascade size distribution roughly follows a power-law with the same exponent observed on square lattices. The waiting time distribution is also similar to those observed on square lattices and empirical results~\cite{dobson07}. These suggest that SOC behaviors may also emerge on real powergrid topologies following our model dynamics.

To further examine the robustness of the SOC behaviors in the IEEE 118-bus system, we re-allocate the power stations randomly and examine the system behaviors. As shown in \fig{fig_realdata}(a), the distribution of failures still roughly follows a power-law but large failures are more likely. These results show that powergrid failures are more strongly dependent on the location of power stations than the topology of the system. With an appropriate allocation of power stations, large failures are more suppressed since the original load from a broken link can be shared evenly to the other links. Unlike the sandpile model composed of all identical nodes, power stations in powergrids play a crucial role and their locations impact the system behaviors. Nevertheless, regardless of the location of the power stations, the system still shows a general picture consistent with SOC.

In addition to the simulations on a real powergrid topology, we also examine the spatial-temporal rescaling using the empirical data of waiting times, obtained from the data spanning 15 years of power outages across the United State ~\cite{realdata}. Since the data do not show the failure of individual links in a blackout, we cannot identify blackouts according to partitions as in~\cite{bak02}  (see Appendix B), but the rescaling according to \req{rescale_rate} by the average failure rate is feasible~\cite{corral04}. As shown in \fig{fig_realdata}(b), the rescaled waiting time distributions with different cutoff size $S$ overlap on a common function, similar to those observed in simulations in \fig{fig_rescale}(c). This again suggests that the empirical results are consistent with the simulated results, and SOC is a potential mechanism underlying real powergrids.

\section{Escaping SOC by proactive maintenance}

If a powergrid is in SOC, large blackouts and their negative impacts~\cite{anderson12, anderson03} are inevitable. As we have seen, the system does not escape form SOC by repairing the broken links after failures, instead this mechanism leads to criticality. Compared with the non-controllable earth crust movement underlying earthquakes, powergrids are composed of controllable components, and therefore, to avoid large blackouts, one can implement microscopic interventions to drive the system away from SOC. In addition to the remedial repairs, we will explore three proactive maintenance approaches to increase link capacity in advance of large failures:
\begin{enumerate}
\item
\emph{Maintenance based on the global load factor} - at each step without failure, given the average network load factor $F$ is larger than a threshold $\psi$, i.e. $F>\psi$ (see \req{eq_loadF}), the capacity of the link with the highest individual load factor, i.e. the link $(ij)=\argmax_{(ij)}[F_{(ij)}]$, is upgraded by $C_{ij}'(t) = C_{ij}(t) (1+\alpha)$; the above procedure is repeated within a time step until $F<\psi$;
\item
\emph{Maintenance based on local load factors} - at each step without failure, we increase the capacity of all links $(ij)$ if their load factor exceeds the threshold $\psi$, i.e. to upgrade $C_{ij}'(t) = \frac{I_{ij}(t)}{\psi}$ when $F_{(ij)}>\psi$, so that $F_{(ij)}$ falls to $\psi$; 
\item
\emph{Routine maintenance} - the capacities of the $n$ links with the highest load factor are upgraded by $C_{ij}(t+1) = C_{ij}(t) (1+\alpha)$.
\end{enumerate}

\begin{figure*}
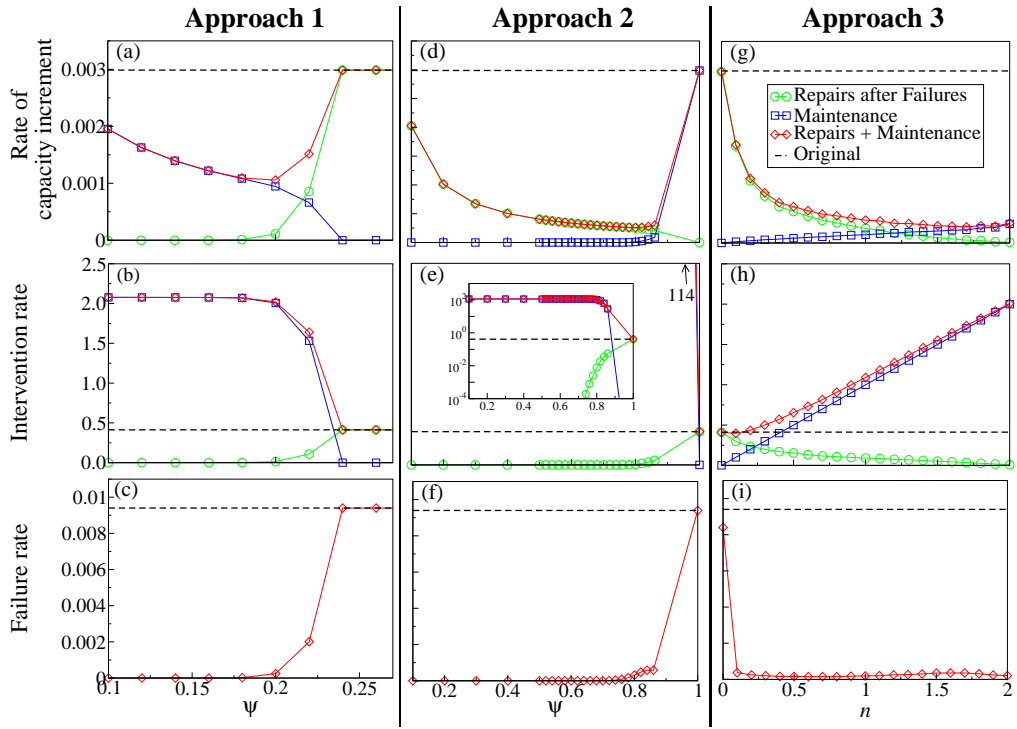

\centering
\begin{tabular}{c|c|c}
\epsfig{figure = M1Resources.eps, width=0.285\linewidth} &
\epsfig{figure = M4Resources.eps, width=0.2215\linewidth}  &
\epsfig{figure = M5Resources.eps, width=0.2215\linewidth} \\
\epsfig{figure = M1Links.eps, width=0.285\linewidth} &
\epsfig{figure = M4Links.eps, width=0.2215\linewidth} &
\epsfig{figure = M5Links.eps, width=0.2215\linewidth} \\
\epsfig{figure = M1FailureRate.eps, width=0.285\linewidth} &
\epsfig{figure = M4FailureRate.eps, width=0.219\linewidth} &
\epsfig{figure = M5FailureRate.eps, width=0.219\linewidth} \\
\end{tabular}
\caption{
(a, d, g) The rate of capacity increment, (b, e, h) the intervention rate (i.e. the average number of maintenance and repairs per step), and (c, f, i) the failure rate of the system, by applying the three maintenance approaches. The results are obtained by averaging 5 realizations of $L\times L$ square lattice with $L=30$ and $N=L^2=900$ nodes, with $5\%$ of the nodes to be power stations. The statistics is obtained for $N\times 10^2$ updates after an equilibration of $N\times 10^4$ updates.
}
\label{fig_approaches}
\end{figure*}

We first discuss the results of the first approach. As shown by the blue ($\Box$) and green ($\bigcirc$) curves respectively in \fig{fig_approaches}(a), for $\psi<0.2$, a large amount of extra capacity is required in proactive maintenance, but only a small amount of extra capacity is required to repair the broken links since blackouts are rare as shown in \fig{fig_approaches}(c); the system escapes from SOC. When $\psi$ increases from $\psi=0.2$ to $\psi=0.25$, less extra capacity is required for maintenance but more capacity is used for repairing the failed links. Beyond $\psi\approx 0.25$, the amount of extra capacity required for proactive maintenance almost vanishes, which indicates that the global load factor $F$ seldom reaches $\psi$ to activate the proactive procedures; this may also mark the typical load factor which triggers the failures. The system returns to SOC and further reduces to the original no-maintenance case when $\psi\rightarrow 1$.

\begin{figure}
\centerline{
\epsfig{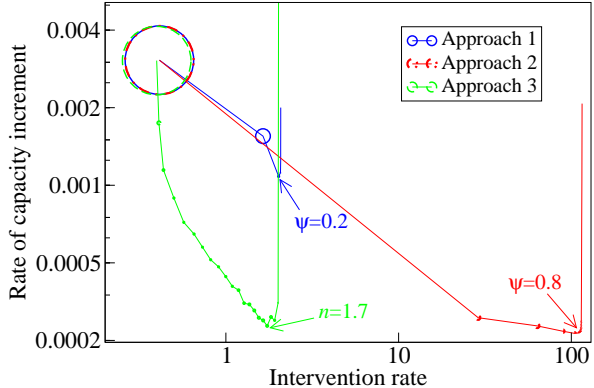}
}
\caption{
The trade-off between the rate of capacity increment and the intervention rate of the three proactive maintenance approaches. The size of the nodes corresponds to the rate of blackouts.
}
\label{fig_tradeoff}
\end{figure}

Interestingly, compared with the original case without maintenance (i.e. $\psi\rightarrow 1$), (i) the total amount of extra capacity required for all interventions (i.e. proactive maintenance or remedial repairs) is less and attains a minimum at $\psi\approx 0.2$, and (ii) at the same time failures are less frequent, as shown in \fig{fig_approaches}(a) and (c) respectively. It is because maintenance leads to an even distribution of redundant capacity, compared with the biased distribution of excessive capacity on individual heavily loaded links after failures.  However, a large number of maintenance is required in this approach as shown in \fig{fig_approaches}(b), resulting in a trade-off between the required amount of extra capacity and the required number of interventions (i.e. the total number of maintenance and repairs) to sustain the system as shown in \fig{fig_tradeoff}.

We then discuss the second and the third maintenance approach. As shown in \fig{fig_approaches}(d) - (f), the second approach behaves similarly as the first approach, except that a much larger number of interventions (mainly maintenance) are required in exchange for a smaller amount of extra required capacity (see also \fig{fig_tradeoff}). On the other hand, as shown in \fig{fig_approaches}(g) and (h), the third approach requires only a small amount of extra capacity achieved by a small number of interventions, outperforming both the first and the second approach. Nevertheless, it is less effective in eliminating failures as shown \fig{fig_approaches}(i) unless $n\gtrsim 3$ at which much more extra capacity are required. A better comparison is shown in \fig{fig_tradeoff}, where the third approach - a simple routine maintenance - is the most cost-effective approach to drive the system away from SOC. In summary, all three approaches suppress blackouts and at the same time use less extra capacity compared with the original case without proactive maintenance.

We remark that powergrids with proactive maintenance may share similarities with the non-conserving variants of SOC models~\cite{olami92, drossel92}, in which the SOC characteristics are lost. While the former extends capacity in advance to suppress blackouts, the latter dismisses a fraction of energy during toppling to suppress avalanches, leading to subcritical behaviors in both kinds of systems~\cite{broker97, pruessner02}. As a result, proactive maintenance are effective to suppress large failures since they drive the system away from SOC.

\section{Discussion}

Powergrids were suggested to be in self-organized criticality (SOC). To verify the conjecture, we introduced a model of powergrids which constantly evolves and upgrades to satisfy the increasing energy demand, resulting in a dynamics of intermitten failures and recovery. The model demonstrates various SOC characteristics and an analogy to earthquakes. Some of these findings are verified by implementation on real powergrid topology and empirical data. The model behaviors and the data analyses provide further evidences to suggest that an evolving powergrid is in SOC.

If powergrids are in SOC, what are the consequences? Large blackouts are inevitable and unpredictable. Since they have a great negative impact, it is essential to suppress them, and the way to achieve the goal is to drive the system away from SOC, or otherwise it self-organizes to criticality again. Unlike earthquakes which involve the non-controllable earth crust, powergrids are composed of controllable components; intervention may greatly increase the predictability and controllability of blackouts. Several proactive maintenance approaches are suggested in additional to the ordinary remedial repairs, and are  shown to be effective in suppressing SOC and large blackouts. While the present study focuses on powergrids, its strong connection with SOC would make the results relevant to other SOC systems, especially in terms of controllability.

\section{Acknowledgement}

CHY acknowledges the Dean's Research Fund 04115 ECR-5 2015 of the Hong Kong Institute of Education and the Research Grants Council of Hong Kong (ECS Grants No. 28300215). AZ is supported by the National Natural Science Foundation of China (Grant No. 11547188) and the Young Scholar Program of Beijing Normal University(Grant No. 2014NT38). KYMW is supported by a grant from the Research Grants Council of Hong Kong (grant number 605813).\\


The author declares no competing financial interests.

\section*{APPENDIX A: Detrended fluctuation analysis}

We employ the detrended fluctuation analysis (DFA) \cite{hu01, lennartz08} to examine the long term memory, i.e. persistence or anti-persistence, of the sequence of waiting time between blackouts, and the size of the blackouts. Given a time series $x(i)$, with $i=1,2,\cdots,N$, we first define the cumulative sequence $y_i$ to be
\begin{equation}
y(i)=\sum_{j=1}^i(x(j)-\left<x\right>),
\end{equation}
where $i=1,2,\cdots,n$, and $\left<x\right>$ is the expected value for the whole sequence $x(i)$, given by
\begin{equation}
\left<x\right> = \frac{1}{N}\sum_{i=1}^{N}x(i).
\end{equation}
We then divide the whole sequence into $\lfloor\frac{N}{n}\rfloor$ time window of equal size $n$. In each time window, we best-fit the cumulative sequence in the window with a $l$-th order polynomial function $y_{fit}(i)$ given by
\begin{equation}
y_{fit}(i) = \sum_{k=0}^l\beta_ki^k
\end{equation}
where $\beta_k$'s are the coefficient of the polynomial, and $y_{fit}(i)$ is thus the local trend in the window. We call the analyses $l$-DFA given an $l$-th order polynomial is used as the fit.

We then compute the detrended fluctuation function $\cF(n)$ for time window length $n$, by detrending the cumulative sequence, i.e. subtracting $y(t)$ in each window by the corresponding local trend $y_{fit}(i)$, i.e.
\begin{equation}
z(i) = y(i) - y_{fit}(i).
\end{equation}
For each window length $n$ , we calculate the root-mean-square deviation from the trend as the detrended fluctuation function $\cF(n)$, given by
\begin{equation}
\cF(n)=\sqrt{\frac {1}{n}\sum_{i=1}^n(z(i))^2}.
\end{equation}
We denote the average of $\cF(n)$ over the different time window to be $\avg{\cF(n)}$. By repeating the above processes for different values of $n=2^k$, with $k=1,\dots$, we obtain a relation between $\avg{\cF(n)}$ and $n$. For a power-law relation $\avg{\cF(n)}\propto n^D$, the sequence is random if $D=0.5$, anti-persistent if $D<0.5$, and persistent if $D>0.5$.

To examine the predictability of blackouts, we performed 0-DFA on the sequences of waiting times and the size of blackouts, and compare these results to the corresponding sequences after random reshuffling, i.e. a random re-order of the individual entry of the sequences. If there are no pattern on the sequences, the exponent $D$ should be consistent with those of the reshuffled sequences, and the waiting times and the size of blackouts are unpredictable. On the other hand, if the the exponent $D$ differs from that of the reshuffled sequence, then there are predictability in the sequences, i.e. either the sequences are persistent or anti-persistence. We remark that 1-DFA and 2-DFA were also performed and the results obtained are similar to those of 0-DFA, and thus we only present the results from 0-DFA.

\begin{figure}
\centerline{
\epsfig{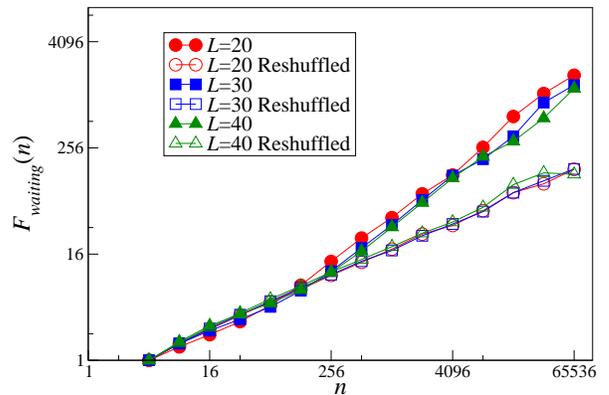}
}
\caption{
$\cF_{\text{waiting}}(n)$ as a function of time window length $n$, for the original and the reshuffled sequence of waiting times for different system size $L$. 
}
\label{fig_DFAInter}
\end{figure}

Figure S\ref{fig_DFAInter} shows $\cF_{waiting}(n)$ as a function of $n$ obtained by 0-DFA for the sequence of the waiting times. Interestingly, $\cF_{waiting}(n)$ seems to be characterized by two different power-laws. When $n$ is small ($n=2^2$ to $2^5$ for $L=20$, $n=2^2$ to $2^6$ for $L=30$ and $n=2^2$ to $2^{7}$ for $L=40$), $D=0.48, 0.49, 0.52$ for $L=20, 30, 40$ respectively, consistent with the corresponding reshuffled sequences. This implies that for a short sequence of blackouts, the waiting times are random and unpredictable. However, for a longer sequence of waiting times ($n=2^6$ to $2^{16}$ for $L=20$,  $n=2^7$ to $2^{16}$ for $L=30$ and $n=2^{8}$ to $2^{16}$ for $L=40$), $D$ deviates from those of the reshuffled sequences and become $0.88, 0.86, 0.85$ for $L=20, 30, 40$ respectively. This suggests that persistence exists if we consider a long sequence of blackouts, i.e. there is a long waiting time which may be a consequence of a large blackout given a sufficiently long time. Moreover, the exponent $D$ changes abruptly, separating an unpredictable range ($D\approx 0.5$) from a predictable range ($D>0.5$). As we can see from \fig{fig_DFAInter}, the larger the system size, the larger the values of $n$ where the system remains unpredictable, implying a lower predictability in larger systems.

\begin{figure}
\centerline{
\epsfig{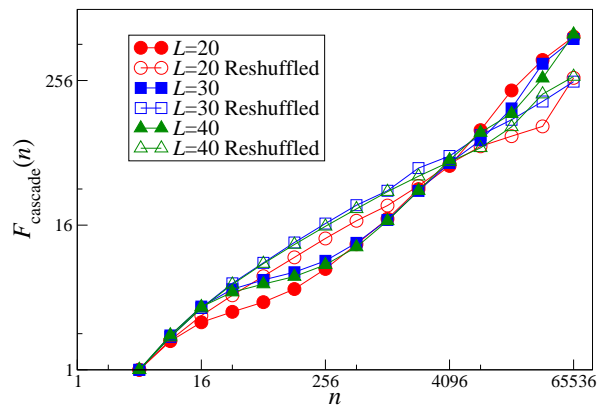}
}
\caption{
$\cF_{\text{cascade}}(n)$ as a function of time window length $n$, for the original and the reshuffled sequence of blackout size for different system size $L$. 
}
\label{fig_DFACas}
\end{figure}

Similar to waiting times, we performed 0-DFA on the sequence of blackout sizes. As shown in \fig{fig_DFACas}, $\cF_{cascade}(n)$ is consistent with the corresponding reshuffled sequences at small $n$  ($n=2^2$ to $2^3$ for $L=20$, $n=2^2$ to $2^4$ for $L=30$ and $L=40$), where $D=0.79, 0.87, 0.87$ for $L=20, 30, 40$ respectively. This implies a persistent behavior in the sequence of blackout size, which may be a consequence of the memory in the loading factor, i.e. the network loading increases gentlely which leads to consecutive blackouts of similar size. We remark that such persistence is also observed for the reshuffled sequence, since small blackouts are much more common (see \fig{fig_sizeDis}). For the intermediate range of $n$ ($n=2^4$ to $2^{7}$ for $L=20$,  $n=2^5$ to $2^{8}$ for $L=30$ and $L=40$), $D=0.30, 0.26, 0.25$ for $L=20,30,40$ respectively, suggesting a slight anti-persistence. This may come from the cycle of blackouts, where large blackouts are usually followed by a period of small blackouts, due to the extensive repairing after a large blackout. Finally, at large $n$ ($n=2^8$ to $2^{16}$ for $L=20$,  $n=2^9$ to $2^{16}$ for $L=30$ and $L=40$), $D$ increases again and becomes $0.83, 0.82, 0.81$ respectively for $L=20, 30, 40$, suggesting the occurrence of a large blackout given a sufficiently long time. Finally, the larger the system size, the larger the values of $n$ where the system remains unpredictable, similar to that observed for waiting times.

\section*{APPENDIX B: Partition for spatio-temporal rescaling of waiting time distribution}

\begin{figure}[h]
\centerline{
\epsfig{figure=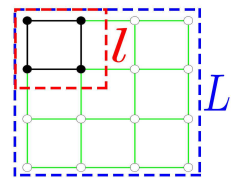, width=0.35\linewidth}
}
\caption{A partition of size $l\times l$ in a network with size $L\times L$.}
\label{fig_partition}
\end{figure}

To examine the spatio-temporal relation of the occurrence of blackouts in our model, we follow~\cite{bak02} to rescale the waiting time distribution. We first group the nodes in the network into square partitions of size $l\times l$, as shown in \fig{fig_partition}. We then define the \emph{constituent links} of a partition to be those with both terminal nodes located inside the partition. Since there is no epicentre in blackouts, we consider a failure has occurred in a partition when at least one of its constituent links breaks; the size of the failure is defined by the number of failed links within the partition. The waiting time in a partition is defined to be the time between consecutive failures in the partition. We further define a cutoff size $S$ such that only the waiting time between failures with size greater than or equal to $S$ is considered. Finally, we denote the waiting time distribution in a $l\times l$ partition with a cutoff size $S$ to be $P_{S,l}(T)$, such that $P_{S,l}(T)$ with various $S$ and $l$ collapse onto the same function given by \req{eq_collapse}. The original and the rescaled distributions are shown in \fig{fig_rescale} (a) and (b).

\section*{APPENDIX C: Waiting time as a function of Redundancy ratio}

\begin{figure}[h]
\centerline{
\epsfig{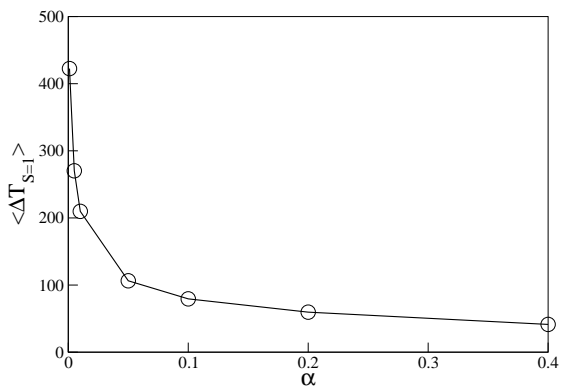}
}
\caption{
The average waiting time as a function of $\alpha$, obtained on $L\times L$ square lattices with $L=30$ and $N=L^2=900$ nodes, with $r=45$ of the nodes to be power station. The statistics is obtained for $1\times 10^6$ updates after an equilibration of $1\times 10^5$ updates.
}
\label{fig_AvgInterTime}
\end{figure}

\begin{figure}[h!]
\centerline{
\epsfig{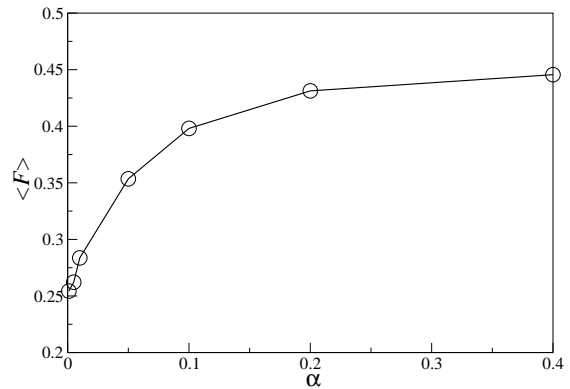}
}
\caption{
The expected value of the average network load factor as a function of $\alpha$, obtained on $L\times L$ square lattices with $L=30$ and $N=L^2=900$ nodes, with $r=45$ of the nodes to be power station. The statistics is obtained for $N\times 10^2$ updates after an equilibration of $N\times 10^4$ updates.
}
\label{fig_AVGF}
\end{figure}

We will discuss the waiting time between two failures as a function of the redundancy ratio $\alpha$. As shown in \fig{fig_AvgInterTime}, the average waiting time starts at a high value at the smallest $\alpha$, decreases with $\alpha$, suggesting that the failure rate decreases with $\alpha$. As shown in \fig{fig_AVGF}, the average network load factor $\avg{F}$ (see \req{eq_loadF}) increases as $\alpha$ increases, in a trend opposite to that of the average waiting time. These results imply that at the smallest $\alpha$ value, failures are less frequent since the demand increments are more gentle and the average network load factor $\avg{F}$ stays at a low value for a large amount of time. Increasing $\alpha$ in this regime would increase $\avg{F}$, which leads to an increase of failure rate, and hence a decrease in the average waiting time.



\end{document}